\documentclass[11pt,fleqn]{article}   
\usepackage{amssymb,cite,amsmath}

\catcode`\@=11
\@addtoreset{equation}{section}
\def\theequation{\arabic{section}.\arabic{equation}}
\def\appendix{\renewcommand{\thesection}{\Alph{section}}\setcounter{section}{0}
              \renewcommand{\theequation}
            {\mbox{\Alph{section}.\arabic{equation}}}\setcounter{equation}{0}}

\def\maketitle{\thispagestyle{empty}\setcounter{page}0\newpage
                \renewcommand{\thefootnote}{\arabic{footnote}}
                  \setcounter{footnote}0}
\renewcommand{\thanks}[1]{\renewcommand{\thefootnote}{\fnsymbol{footnote}}
               \footnote{#1}\renewcommand{\thefootnote}{\arabic{footnote}}}

\renewcommand{\title}[1]{\begin{center}\Large\bf #1\end{center}\rm\par\bigskip}
\renewcommand{\author}[1]{\begin{center}\Large #1\end{center}}
\newcommand{\address}[1]{\begin{center}\large #1\end{center}}

\def\dinfn{\smallskip Dipartimento di Fisica, Universit\`a di Trento\\ 
                           and Istituto Nazionale di Fisica Nucleare,\\
                                   Gruppo Collegato di Trento, Italia}

\def\Idinfn{\address{\dinfn}}

\newcommand{\email}[1]{e-mail: \sl #1@science.unitn.it\rm}

\newcommand{\femail}[1]{\thanks{\email{#1}}}

\def\babs{\hrule\par\begin{description}\item{Abstract: }\it} 
\def\eabs{\par\end{description}\hrule\par\medskip\rm}
\renewcommand{\date}[1]{\par\bigskip\par\sl\hfill #1\par\medskip\par\rm}
 
\newcommand{\s}[1]{\section{#1}}

\def\hs{\qquad}               
\def\nn{\nonumber}            
\def\beq{\begin{eqnarray}}    
\def\eeq{\end{eqnarray}}      
\def\at{\left(}               
\def\ct{\right)}              
\def\R{{\hbox{{\rm I}\kern-.2em\hbox{\rm R}}}}   
\def\H{{\hbox{{\rm I}\kern-.2em\hbox{\rm H}}}}   
\def\N{{\hbox{{\rm I}\kern-.2em\hbox{\rm N}}}}   
\def\C{{\ \hbox{{\rm I}\kern-.6em\hbox{\bf C}}}} 
\def\Z{{\hbox{{\rm Z}\kern-.4em\hbox{\rm Z}}}}   
\def\ii{\infty}                                  

\def\be{\beta}
\def\ga{\gamma}

\def\om{\omega}

\def\De{\Delta}
\def\La{\Lambda}

\def\Om{\Omega}

\begin{document}

 
\title{Hawking Radiation as Tunneling for Extremal and Rotating
  Black Holes} 
\author{Marco Angheben\femail{angheben}, Mario Nadalini\femail{nadalini}, 
Luciano Vanzo\femail{vanzo} and \\
Sergio Zerbini\femail{zerbini}}
\Idinfn

\babs
The issue concerning semi-classical methods recently developed in deriving 
the conditions for Hawking radiation as tunneling, is revisited and
applied also to rotating black hole solutions as well as to
the extremal cases. It is noticed how the tunneling method fixes the
temperature of extremal black hole to be zero, unlike the Euclidean
regularity method that allows an arbitrary compactification period. 
A comparison with other approaches is presented. 
\eabs

\s{Introduction}

Although several derivations of the Hawking radiation \cite{hawk}
have been proposed in the literature, mostly relying on quantum field
theory on a fixed background (see the general references
\cite{birrel,fulling,novikov,wald}), it 
is interesting that only recently a description of black
hole radiation and back reaction as a tunneling effect has been
investigated semi-classically 
\cite{kraus94,kraus95,Keski-Vakkuri:1996gn,pad99,parikh00,pad02,Parikh:2002qh}.   
The essence for such calculations is the computation of radial
trajectories\footnote{The bulk of black holes emission is mainly
  contained in s-waves.} 
in the static or stationary region representing the 
domain of outer communication of the black hole, except that an
infinitesimal region behind the event horizon is allowed, which in
fact plays a major role. This is because the causal structure
of the horizon (it is  
a null, future directed hyper-surface) makes it impossible to
travel from inside to infinity along classically permitted
trajectories. This is one way to understand the frequently  
encountered claim
that the horizon radiation must originate from just outside the
event horizon itself. In that case it would be quite difficult to
detect the 
imaginary part of the black hole free energy in the effective action
formalism, unless one is willing to give up its locality.

In the tunneling approach instead, the particles
are allowed to follow classically forbidden trajectories, by starting
just behind the horizon onward to infinity. The particles must then
travel necessarily back in time, since the horizon is locally to the
future of the static or stationary external region. The classical,
one-particle action becomes complex, signaling 
the classical impossibility of the motion, and gives the amplitude an
imaginary part which may be considered as a ``first quantization''
transition amplitude, or equivalently a free field amplitude, knowing 
that the classical action provides a semi-classical approximation to
free field propagators \footnote{Investigations of black
  holes from the point of view of interior degrees of freedom are
  discussed in \cite{Barvinsky}}. We stress the fundamentally
different physics governing ingoing particles. These can fall behind the
horizon along classically permitted trajectories, with a capture cross
section of the order of the horizon area. Hence the action for ingoing
particles must be real; this will be an important point in our
description of black hole radiation.

Given a generic metric representing such a non rotating black hole
solution in the static coordinates (Schwarzschild gauge), we have
\beq
ds^{2} = -A(r)dt^2 + B^{-1}(r)dr^2+ C(r)h_{ij}dx^{i}dx^{j}\,,
\label{BHS}
\eeq
where the coordinates are labeled as $x^{\mu} = (t,r,x^{i}),
(i=1,...,D)$. 
The metric  $h_{ij}$ is a function of the coordinates $x^{i}$
only, and we shall refer to this metric as the horizon metric. To
insure the black hole has finite entropy, we take
the horizon to be a compact orientable manifold, say $\cal M$. Black hole  
solutions are defined by  functions $A(r)$ and $B(r)$  
having simple and positive zeroes. This is only a necessary condition
to have a black hole; we must also require that the domain of outer
communication be ``outside of the black hole'', i.e. it should correspond to
values of the radial coordinate larger than the horizon and extending
up to spatial infinity. In most cases $B(r)=A(r)$, and $C(r)=r^2$, but  
in order to treat more general coordinate system (e.g. isotropic ones)
and dilatonic black holes (black holes interacting with scalar
fields), we shall also consider $B(r)$ different 
from $A(r)$ and possibly $C(r)$ not equal to $r^2$. Interesting black
holes with such metrics can be obtained in the
Einstein-Maxwell-dilaton coupled system. An example illustrating the first
feature is the following two-parameter family solution 
\beq
ds^2=
-\left(1-\frac{r_+}{r}\right)dt^2+\left(1-\frac{r_+}{r}\right)^{-1}
\left(1-\frac{r_-}{r}\right)^{-1}dr^2+r^2dS_2^2
\eeq
where $dS_2^2=d\theta^2+\sin^2\theta d\phi^2$ is the round metric on
the sphere.
The dilaton is $\exp(2(\phi-\phi_{\ii}))=(1-r_-/r)^{-1/2}$; the hole has
magnetic charge $q_m=3r_+r_-/16$, horizon radius $r_+=2m$ (which
defines $m$) and can be
extended to a non singular, geodesically complete solution with
horizons and asymptotically flat infinities. The near horizon geometry
in the near extremal limit $r_+=r_-$ is determined by a
two-dimensional black hole of the Jackiw-Teitelboim model, and its
entropy can be accounted for exactly by two-dimensional dilaton
gravity. As an example illustrating the second feature, we propose the
following Kaluza-Klein $4D$ black hole 
\beq
ds^2=-\Delta^{-1/2}(1-2mr^{-1})dt^2+\Delta^{1/2}(1-2mr^{-1})^{-1}dr^2+
\Delta^{1/2}r^2dS_2^2
\eeq
where $\De=1+2mr^{-1}\sinh^2\ga$ and $\ga$ is a real constant. 
The metric is asymptotically flat with an event horizon at $r=2m$. The
dilaton is $\exp(-4\phi/\sqrt{3})=\Delta$ and there is an electric field
corresponding to a charge $Q=m\sinh2\ga/2$.  This metric and dilaton
solve the Einstein-Maxwell-dilaton field equations and can be
obtained from dimensional reduction of Kaluza-Klein theory, i.e. by
finding a solution in five dimensions and then putting the $5D$ metric
in the form 
\[
ds^2=e^{-4\phi/\sqrt{3}}(dx_5+2A_adx^a)^2+e^{2\phi/\sqrt{3}}g_{ab}dx^adx^b
\]
Then $g_{ab}$, $\phi$ and $A_a$ are the sought for $4D$ metric,
dilaton and gauge field, respectively.

When the cosmological constant is non positive, often one has only
one horizon defined by the largest positive root of the lapse
function, and the range of $r$ is an infinite interval. An exception
are the topological black holes with higher genus
\cite{vanzo,brill,mann}, that have a Cauchy 
horizon hidden behind the event horizon. When the
cosmological constant is positive, there exists the possibility of
multiple horizons, some related to black holes, some others to
infinity (cosmological horizons) and the range of $r$ is a finite
interval.\\  
The form of the metric and choice
of coordinates as given above has always been the natural setting to
describe Hawking radiation. However, in the tunneling approach of
Wilczek and co-workers, the system of coordinates introduced
by Painlev\'e \cite{pain} plays a special role.
The Painlev\'e coordinates are associated with a redefinition of the
coordinate time, and are given by \cite{pain,ve03}
\beq
T=t \pm \int dr\sqrt{\frac{1-B(r)}{A(r)B(r)}}\,.
\label{pain}
\eeq 
As a result, the  BH metric becomes a stationary one. In fact a
simple computation leads to 
\beq
ds^{2} = -A(r)dT^2 \pm 2\sqrt{(1- B(r))\frac{A(r)}{B(r)}}dr dT+ dr^2+ r^2
h_{ij}dx^{i}dx^{j}\,.
\label{BHP}
\eeq
Applying the tunneling method  to several BH solutions
\cite{parikh00,med,Medved:2001ca,ve03} it is found that the tunneling
probability at the leading order, for a particle to escape, is given
by  
\beq
\Gamma \equiv e^{-\beta_H E}\,,
\label{W}
\eeq
where $E$ is the energy of the emitted particle, assumed to be
sufficiently small with  respect to the total energy of the BH. Of
course the flux at infinity will be reduced by the corresponding gray
body factors. However, the thermal nature of the flux can be inferred
only because the coefficient $\be_H$ is known to be the inverse
temperature of the hole, so that \eqref{W} can really be interpreted
as a Boltzmann distribution.\\
Alternatively, the so called method of complex paths has been
introduced in several papers \cite{pad99,pad02} (a
review can be found in \cite{Padmanabhan:2003gd}). Here the correct
result has been obtained referring to the seminal paper of Hartle and
Hawking \cite{hw76} and making use of several  coordinates, including the 
Schwarzschild ones,  but at the expense of the introduction of the
concept of multiple mapping, necessary for recovering the
covariance. This is because under changes of coordinates within a time
slice the results were not invariant.\\
In this paper, we would like to show that it is possible to work in
the Schwarzschild gauge, making use of a variant of the tunneling
method, recovering the general covariance simply by observing that a correct
use of the theory of distributions in curved (static) space-time
requires the introduction of the proper spatial distance, as defined
by the spatial metric
\beq
d\sigma^2=\frac{d^2r}{B(r)}+C(r)h_{ij}dx^{i}dx^{j}\,, 
\label{inv}
\eeq
a quantity that is {\it invariant} under redefinitions of the spatial
coordinate system, and more generally with respect to the subgroup of 
the gauge group consisting of time recalibration and spatial
diffeomorphisms, defined by \cite{cattaneo,zelmanov,landau,moller72} 
\begin{eqnarray}
& & t'=t'(t,r,x^i)\\
& & r'=r'(r,x^i)\\
& & x^{j'}=x^{j'}(r,x^i)\,.
\label{inv1}
\end{eqnarray}
As a result, it will not be necessary to work with a complex action for
ingoing particles.

The remainder of this paper is organized as follow: 

In Section II we show how an
ambiguity arises if one insists to treat coordinate singularities as 
defining distributions, an how it can be resolved by transition to invariant
quantities. The role of ingoing particles is also briefly discussed.
In Sections III and IV, we extend the results to a class of extremal
black holes and cosmological horizons, respectively. The tunneling
method makes it easy to derive the fundamental property of extremal
black holes, the vanishing of their temperature. It is worth noticing
that the absence of conical singularities in the Wick rotated solution
allows for arbitrary compactification of Euclidean time. We consider
the extension of the method to rotating cases in Section V, 
especially the three-dimensional BTZ
black hole and the Kerr-anti-de Sitter (KAdS) black hole in four
dimensions, which also covers the Kerr solution in flat space. Some 
conclusion and observations are reported in Section VI.

\section{Hawking radiation as tunneling: single horizon}

In this Section, in order to illustrate our approach, we will
elaborate a variant of the tunneling method
in the static case of a single horizon. The multiple horizons case as
well as the rotating case will be considered in the next Sections.

To begin with, recall the metric we are interested in reads
\beq
ds^{2} = -A(r)dt^2 + B^{-1}(r)dr^2+ C(r)
h_{ij}dx^{i}dx^{j}\,.
\label{ansatz}
\eeq
We shall consider a scalar particle moving in this classical BH
background. Within the semi-classical approximation and being
interested only in the leading contribution, we may neglect particle
self-gravitation. Thus, the relevant quantity is the classical action
$I$, which satisfies the relativistic Hamilton-Jacobi equation
\beq
g^{\mu \nu}\partial_\mu I \partial_\nu I+m^2=0\,,
\label{m1}
\eeq
which reads
\beq
-\frac{1}{A(r)}(\partial_t I)^2+B(r)(\partial_r I)^2+
\frac{1}{C(r)}g^{i j}\partial_i I \partial_j I+m^2=0\,.
\label{m2}
\eeq
As usual, due to the symmetries of the metric, one is looking for a
solution in the form 
\beq
I=-E t+W(r)+J(x^i)\,.
\eeq
As a consequence
\beq
\partial_t I=-E\,,\,\,\,\ \partial_r I=W(r)\,,\,\,\,\partial_i I=J_i\,,
\eeq
where $J_i$ are constants (some of them may be chosen to be zero).
Thus, the classical action is given by (the sign of the second term
corresponds to an outgoing particle, it would be opposite for an
ingoing particle)
\beq
I=-Et+\int  \frac{dr}{\sqrt{A(r)B(r)}}\sqrt{ E^2-A(r)\left(m^2+
\frac{g^{i j}J_i J_j}{C(r)}\right)}+J(x_i)\,. 
\label{m3}
\eeq

It is important to recognize that the action for ingoing particles has
to be real, since a particle can fall down in a black hole along a
classically permitted trajectory, if only the impact parameter is
within the order of magnitude of the horizon radius. Hence the
apparent singularity of \eqref{m3} for ingoing particles near the
horizon is spurious, and should be eliminated by transition to a
coordinate system which is regular on the horizon.\footnote{For a
 somewhat different treatment, see \cite{Shankaranarayanan:2000gb}.}

Now one has to face with the key point. If we are dealing with a BH
solution, the contribution due to the integral over the radial
coordinate is divergent as  
soon as the integration includes the horizon. One needs a regularization, the 
natural one is the equivalent of the Feynman prescription and it consists in 
deforming the contour and, as it is well known, this
produces an imaginary part, whose physical consequence is associated
with the tunneling process. However, this naive approach leads to an 
imaginary contribution which is one half the correct one. For example,
in the case 
of the 4-dimensional Schwarzschild BH, $A(r)=B(r)=1-\frac{2MG}{r}$,
$C(r)=r^2$ and $r_H=2MG$, a direct computation leads to
\beq
\mbox{Im}I=\mbox{Im}W= \pi r_H E\,.
\eeq
 
Furthermore, if one repeats the above calculation making use of
another coordinates system, for example the isotropic ones, defined
by (here we consider the case $A(r)=B(r)\,, \,\, C(r)=r^2$)
\beq
t \rightarrow t\,,\,\,\,
r \rightarrow \rho\,,\,\,\,\, \ln {\rho}=\int \frac{dr}{r\sqrt{A(r)}}\,,
\eeq
the metric assumes the form 
\beq
ds^2=-A(r(\rho))dt^2+\frac{r^2(\rho)}{\rho^2}\left(d\rho^2+\rho^2
h_{ij}dx^{i}dx^{j}\right)\,.
\label{ansatz1} 
\eeq
In this system of coordinates, the spatial metric is no longer
singular at the horizon, and, in the new radial coordinate $\rho$, has
the general form  
(\ref{ansatz}).
For the 4-dimensional Schwarzschild case, the metric in this
coordinates is well known and reads
\beq
ds^2=-dt^2
\frac{\left(1-\frac{r_H}{4\rho}\right)^2}{\left(1+\frac{r_H}{4\rho}\right)^2}
+\left(1+\frac{r_H}{4\rho}\right)^4(d\rho^2+\rho^2 dS^2_2)\,.
\eeq
This form of the metric is still static, but with a radial part
regular at the horizon $\rho=r_H$. We may apply again Eq. (\ref{m3})
deforming the contour and a direct computation gives the correct
result 
\beq
\mbox{Im}I=\mbox{Im}W= 2\pi r_H E\,.
\eeq
 
The reason of this discrepancy can be understood observing that in a
curved manifold, the non locally integrable function $\frac{1}{r}$ 
does not leads to a covariant  distribution  $\frac{1}{r\pm i0}$. One
has to make use of the invariant distance defined by Eq. (\ref{inv}). 

If we limit ourselves to the s-wave contribution, only the  radial
part enters the game and we have for the relevant contribution of the
classical action 
\beq
W(\sigma)=\int  \frac{d \sigma}{\sqrt{A(r(\sigma))}}
\sqrt{ E^2-A(r(\sigma))m^2 }\,. 
\label{m4}
\eeq
We may treat the case $A(r)$ different from $B(r)$, but both vanishing
at $r=r_H$. 
Since the leading contribution is coming from the horizon, we may
use the following near-horizon approximation, 
\beq
A(r)=A'(r_H)(r-r_H)+...\,,\,\,\,\,B(r)=B'(r_H)(r-r_H)+...\,,
\eeq
\beq
\sigma =\int \frac{dr}{\sqrt{B(r)}}=\frac{2}{\sqrt{B'(r_H)}}\sqrt{r-r_H}\,.
\eeq
Thus, we have the invariant result
\beq
W(\sigma)=\frac{2}{\sqrt{A'(r_H)B'(r_H)}} \int  \frac{d \sigma}{\sigma}
\sqrt{ E^2-A(r(\sigma))m^2 }\,. 
\label{m5}
\eeq
The integral is still divergent as soon as $\sigma \rightarrow 0$,
namely the horizon is reached, but now the  prescription corresponding
to the Feynman propagator selects the correct imaginary part for the
classical action 
\beq
I=\frac{2\pi i}{\sqrt{A'(r_H)B'(r_H)}}\,E+ \mbox{(real contribution)}\,.
\label{m6}
\eeq
For example, in the  4-dimensional Schwarzschild  case, one gets
\beq
I= 2 i \pi r_H E+ \mbox{(real contribution)}\,.
\label{m61}
\eeq

and the semi-classical emission rate, with only the leading
term linear in $E$  included, reads 
\beq
\Gamma \equiv e^{-2 \mbox{Im}I}= e^{-\frac{4\pi E}{\sqrt{A'(r_H)B'(r_H)}}}\,.
\label{cc}
\eeq
This turns out to coincide with the standard Boltzmann factor as soon
as one recognizes that  
\beq
\beta_H=\frac{4\pi}{\sqrt{A'(r_H)B'(r_H)}}\,.
\eeq 
is the Hawking temperature measured at the infinity for a
generic asymptotically flat BH. The same conclusion is also valid for 
asymptotically anti-de Sitter (AdS) BHs. More attention deserves the
asymptotically de Sitter (dS) case, where multiple horizons can be
present. We would like to note that $\beta_H$ can be interpreted as
the period of Euclidean time in the black hole Euclidean section,
fixed by the regularity requirement of the absence of a conical
singularity at the horizon.

\section{Hawking radiation as tunneling: the extremal case}

The physics of extremal black holes has many ramifications extending 
from classical black hole thermodynamics to string theories, 
mirrored by the presence of several, radically different
solutions. In the following section therefore, we 
will not cover all the extremal black hole solutions that 
exist in theories involving coupled gravitational, electromagnetic
and scalar fields \cite{maeda,GHS}, since for certain values of the
dilaton coupling (the term $\exp(-2a\phi)F^2$ in the Lagrangian), the
thermodynamic description breaks down
\cite{Holzhey:1991bx,Preskill:1991tb}. To give
an illustration of the method regarding  extremal black holes (those
having vanishing surface gravity), we consider instead some specific
examples, starting with the GHS  black hole \cite{GHS} defined by the
condition 
\beq
Q^2=2M^2e^{2\phi_0}\,,
\eeq 
where $Q$ is the charge, $M$ a mass and $\phi_0$ is the constant value of the 
dilaton field. The metric reads
\beq
ds^2=-dt^2+\frac{dr^2}{(1-\frac{C}{r})^2}+r^2 dS^2_2\,,
\eeq
where $C=2M e^{-\phi_0}$.
The proper distance from a point of radial coordinates $r_1$ and the
 horizon $r=C$ is 
\beq
\sigma=r-r_1-\ln \frac{r-C}{r_1-C}\,.
\eeq
As well known in several extremal cases, the horizon is at infinite
spatial distance from a generic point, although it can be traversed in
a finite proper time. However, since we  have a
trivial $A(r)=1$, the general formula (\ref{m4}) leads to
\beq
W=\sqrt{E-m^2}\int_0^\infty d \sigma\,,
\eeq
a divergent integral, whose analytic regularization exists, but no
imaginary part is present! The conclusion is that the Hawking
temperature vanishes, as was to be expected since without horizon
redshift and conserved gauge charges there is no thermal Hawking
radiation from the hole. This is in agreement with the results
obtained in Refs. \cite{anderson,moretti,binosi}. See, however \cite{ve01}.

Furthermore, extremal black holes
can still radiate in the charged channel, since they have non zero
charge \cite{Gibbons,Vanzo,Page}\footnote{For a different view and
  conclusions, see \cite{Alvarenga:2003tx}.}. The process is known as
super-radiance and 
is not associated to any definite temperature. The point is that the
tunneling method, as it stands, seems unable to give the
super-radiant emission.  We also note that for dilaton coupling $a>0$
the distance 
of the horizon remains finite; for example this happens in the family
\[
ds^2=-A(r)dt^2+A(r)^{-1}dr^2+r^2B(r)d\om^2
\] 

where 
\[
A(r)=\left(1-\frac{r_0}{r}\right)^{\frac{2}{1+a^2}}, \qquad
B(r)=\left(1-\frac{r_0}{r}\right)^{\frac{2a^2}{1+a^2}}
\]

For $a>1$ even the tortoise coordinate remains finite, so the
classical no-hair theorems (relying on the infinite extent of the
tortoise coordinate) require other considerations for their validity.

An example with infinite horizon distance is the 5-dimensional
topological black hole \cite{vanzo,bir}, whose extremal metric reads
\beq
ds^2=-dt^2 \frac{(r^2-r_H^2)^2}{r^2 l^2}+
\frac{dr^2}{\frac{(r^2-r_H^2)^2}{r^2 l^2}}+r^2dH_3^2\,,
\eeq
where $dH_3^3$ is the metric associated with a compact hyperbolic manifold.
Here we have  non trivial $A(r)=B(r)$. 
The proper distance from a point of radial coordinates $r_1$ and the horizon 
$r=r_H$ is
\beq
\sigma=\frac{l}{2}\left(\ln (r_1^2-r_H^2)-\ln (r^2-r_H^2)\right)\,.
\eeq
Again the horizon is located at an infinite distance. Thus,
\beq
A(\sigma)=\frac{(r_1^2-r_H^2)^2e^{-4\frac{\sigma}{l}}}{l^2 r^2(\sigma)}\,.
\eeq
Here
\beq
r^2(\sigma)=r^2_H+(r_1^2-r_H^2) e^{-2\frac{\sigma}{l}}\,.
\eeq
 
The formula  (\ref{m4}) gives, 
\beq
W=\frac{1}{(r_1^2-r_H^2)}\int_0^\infty d \sigma e^{2\frac{\sigma}{l}}\, 
r(\sigma) \sqrt{E^2-m^2 A(r(\sigma))}\,.
\eeq
As usual, the integral is divergent as soon as one point arrives at the 
horizon and no analytic regularization exists giving an imaginary part.
As a consequence, once more the  Hawking temperature is vanishing.
Finally, for solutions with horizon at finite distance one can
even obtain infinite temperature, showing that a thermodynamic
description can be inadequate.   

It should be noted that this correct conclusion can be obtained only if one
makes use of the proper distance. In fact, it is easy to show that if one uses
the radial coordinate in the discussion of the singular integral, one can find
a prescription leading to a non vanishing imaginary part. 
In fact, in the case of the first example, one would have
\beq
W=\sqrt{E^2-m^2} \int_{r_H}^{r_1} d r \frac{1}{(r-r_H)}\,.
\eeq
Deforming the contour, one gets a non vanishing imaginary
part. 
One arrives at the same conclusion in the second example we have considered,
namely  the 5-dimensional extremal topological black hole. Here one
would have 
\beq
W=\int_{r_H}^{r_1} d r \frac{r^2}{(r^2-r_H^2)(r^2+r_H^2)} 
 \sqrt{E^2-m^2 A(r)}\,.
\eeq
Deforming the contour and  making  use  of the formula
\beq
\int \frac{dx}{x^2 \pm i0}f(x)=P\int\frac{dx}{x^2}\mp i\pi f'(0)
\eeq
one gets again a non vanishing imaginary part.

\section{Hawking radiation as tunneling: multiple horizons}

Very recently the so called method of complex paths has been applied
to the case of BH solution having multiple horizons \cite{ind03}. 
The relevant physical example being $(n+1)$-dimensional
Schwarzschild-de Sitter black hole. It is described by the  static
metric with $-g_{00}=g_{rr}^{-1}=A(r)$, where
\beq
A(r)= 1 -\frac{\omega_{n}M}{r^{n-2 }} - \frac{r^{2}}{l^{2}},
\label{f}
\eeq
where $\omega_n$ is a geometrical factor containing also the
gravitational constant and  $M$ can be considered as the mass of the
black hole.  

This metric is a solution of the vacuum Einstein equations with
positive cosmological constant 
\beq
\La=\frac{n(n-1)}{2l^2}\,.
\eeq
The case $n=3$ has been studied in the  paper \cite{gibb77}.

The geometry of  the horizon corresponds to a metric $h_{ij}$ such that
\beq
R_{ij}(h) = (n-2)\;h_{ij}.
\label{rij}
\eeq
namely one has spherical black hole solutions in the asymptotically de
Sitter space-time, where $M$ is the mass parameter of the black
hole. The lapse  
function $A(r)$ has, at least, two positive simple zeroes, the smaller
$r_H$ defines the black hole event horizon, while the larger $r_C$ 
represents the cosmological event horizon \cite{gibb77}. For  example,
for $n=4$, the five dimensional case, one can explicitly find the two roots 
\beq
r_H&=&\frac{l}{\sqrt 2}\at 1-\sqrt{1-4\omega_4M}\ct \nn \\
r_{C}&=&\frac{l}{\sqrt 2}\at 1+\sqrt{1-4\omega_4M}\ct \,.
\eeq

The background manifold corresponds to $M=0$ and is the De Sitter
space-time, the Euclidean counterpart is isometric to $S^{n+1}$. 

Now let us consider a particle located at the classical allowed  $r$,
namely with $r_H < r< r_C$. We may repeat the argument for the 
calculation of the imaginary part of the related  classical
action. Since we are dealing with two horizons, a particle can be in
between by two mutually exclusive tunneling processes 
associated with the event horizon {\it or} with
the cosmological horizon. These contributions  may be evaluated with 
the method developed in the previous Section. Thus, related to the
same particle, we have 
\beq
\Gamma_H \equiv e^{-2 \mbox{Im} I}= e^{-\frac{4\pi E}{A'(r_H)}}\, ,
\label{ccH}
\eeq
and
\beq
\Gamma_C \equiv e^{-2 \mbox{Im} I}= e^{-\frac{4\pi E}{A'(r_C)}}\,,
\label{ccC}
\eeq
where
\beq
\beta_H = \frac{4 \pi}{A'(r_H)}=\frac{4 \pi l^{2}r_H}{(n-2)l^{2}-nr_H^{2}}.
\eeq
and
\beq
\beta_C = \frac{4 \pi}{A'(r_C)}=\frac{4 \pi l^{2}r_C}{(n-2)l^{2}-nr_C^{2}}.
\eeq
Since $\beta_H >0$, it follows that there exists a critical radius
\beq
r_H < r_0=l \sqrt{\frac{n-2}{n}}\,.
\eeq
at which the two horizons coalesce, and we have the Nariai solution,
which represents the largest black hole one can have in de Sitter
space. The total tunneling rate should be a mixture of the two
contributions 
with a relative weight proportional to the ratio of the gray body
factors of the two horizons. The total flux can probably be described
by a flux at some intermediate temperature. The question whether a
global unique temperature can be attributed to Schwarzschild-de Sitter
space-time has been investigated in
\cite{Lin:1998pj,Padmanabhan:2004tz,Choudhury:2004ph}, where such a
temperature 
is found at the price of orbifolding the Euclidean section. But the
quantum state becomes coordinate dependent. It seems that tunneling,
by treating separately the horizons could not relate the
corresponding temperatures in any significant way.  
 
\section{The rotating case}

In this Section, we shall try to extend the method to some rotating 
stationary black hole solution.   

First, let us consider the simplest case, namely the rotating BTZ 
black hole \cite{BTZ}
\beq
ds^2=-A(r)dt^2+\frac{dr^2}{A(r)}+r^2 \at d\phi-\frac{J}{2r^2} dt \ct^2\,,
\label{BTZ}
\eeq
written here in units where the three-dimensional Newton constant is
$8G=1$, and $l^2$ is related to the negative cosmological constant by 
means of $\Lambda=-l^{-2}$, while the lapse function reads 
\beq
A(r)=\frac{r^2}{l^2}-M+\frac{J^2}{4r^2}\,.
\eeq
Here, $ M$ and $J$ are respectively  the mass and angular momentum of
the black hole.  
When $M^2l^2 >J^2$, one has the event horizon located at
\beq
r_H^2=r_0^2 \at 1+\sqrt{1-\frac{J^2}{M^2l^2}} \ct\,,
\eeq
with $r_0^2=\frac{Ml^2}{2}$.
One also has an inner horizon, given by
\beq
r_I^2=r_0^2 \at 1-\sqrt{1-\frac{J^2}{M^2l^2}} \ct\,.
\eeq
In the near horizon approximation, one again has
\beq
A(r)=A'(r_H)(r-r_H)+...\,,
\eeq
\beq
ds^2=-A(r)dt^2+\frac{dr^2}{A(r)}+r_H^2  d\chi^2\,,
\label{BTZA}
\eeq
in which
\beq
\chi=\phi- \Omega t, \qquad \Omega =\frac{J}{2r_H^2}\,,
\eeq
 $\Omega$ being the angular velocity of the horizon. 
As a consequence, we may apply the result of  Section 2 by writing
$I=-Et+J\phi+W(r)$, then passing to the variable $\chi$ (which
transforms $E$ into $E-\Omega J$) and arrive at

\beq
\Gamma_{BTZ} \equiv e^{-2 \mbox{Im} I}= e^{-\frac{4\pi (E-\Omega
    J)}{A'(r_H)}}\,, 
\label{ccC0}
\eeq 
with $E-\Omega J>0$ (see below) and 
\beq
A'(r_H)=\frac{l^2r_H}{2(r_H^2-r^2_I)}\,.
\eeq
The non rotating case corresponds to $J=0$ and $r_I=0$, and the extreme 
case  to $r_I=r_H$. 

It may be useful to recall that the extreme limit in not unique since,
as a rule, there also exists other extremal limits.
One can arrive at this by the techniques
discussed in several papers. For the rotating case see, for example, 
\cite{calda} and  references therein. One alternative extremal limit
for the BTZ rotating black hole turns out to be
\beq
ds^2=-A(r)dt^2+\frac{dr^2}{A(r)}+ r_0^2\at d\phi-\frac{2r}{r_0l}\,dt \ct^2\,,
\label{BTZEX}
\eeq
\beq
A(r)=\frac{4r^2}{l^2}-\frac{r_0^2}{l^2}\,.
\eeq
The horizon is located at $r_H=\frac{r_0}{2}$. We note the absence of
an ergosphere, since the natural Killing field $\partial_t$ is
time-like everywhere. In fact, the metric has no super-radiant modes
and with a linearly diverging local angular velocity it looks like the
field of a rotating disk \footnote{The study of this metrics may be of
  interest as a case for AdS/CFT correspondence.}. 
Again, in the near-horizon approximation, one gets
\beq
\Gamma_{BTZEX} \equiv e^{-2 \mbox{Im} I}= e^{-\frac{4\pi E}{A'(r_H)}}\,,
\label{ccC1}
\eeq 
with
\beq
A'(r_H)=\frac{4r_0}{l^2}\,. 
\eeq
The temperature is $\be_H=\pi l^2r_0^{-1}$; thus, in contrast with the
former extremal black holes, this one radiates thermally like the more
familiar non extremal states.

Along the same lines, one can consider the Kerr-AdS black hole (KAdS)
in four dimensions \cite{carter:1968}. It is convenient to start from
the Kerr-AdS solution 
written in canonical ADM form, where the lapse and shift functions
(the local angular velocity) are seen explicitly
\beq
ds^2=-A^2(r,\theta)dt^2+\frac{dr^2}{B(r,\theta)}+ 
\frac{\rho^2}{\Delta_\theta} d\theta^2 +
\frac{\Sigma^2 \sin^2 \theta}{\rho^2\Xi^2}\at  d\phi-\omega  dt \ct^2\,,
\label{K}
\eeq
In the following $\La=-3l^{-2}$ is the cosmological constant and
\beq
A^2(r,\theta)=\frac{\rho^2 \Delta_r \Delta_\theta}{\Sigma^2}\,, \hs
B(r,\theta)=\frac{\Delta_r}{\rho^2}\,,\hs \Xi=1-\frac{a^2}{l^2}\,,
\eeq
\beq
\rho^2=r^2+a^2 \cos^2 \theta\,,\hs\Delta_r=(r^2+a^2)(1+\frac{r^2}{l^2})-2mr\,,
\eeq
\beq
\Sigma^2=\Delta_\theta(r^2+a^2)^2-\Delta_r a^2 \sin^2 \theta\,, \hs
\Delta_\theta=1-\frac{a^2}{l^2}\cos^2 \theta\,,
\eeq
\beq
\omega=\frac{a\left(\Delta_\theta (r^2+a^2)-\Delta_r\right)\Xi}{\Sigma^2}\,.
\eeq
The limiting value of $\omega$ as $r\to r_H$  is the angular velocity
of the horizon
\beq\label{anghor}
\Omega=\frac{a\Xi}{r_H^2+a^2}
\eeq

The Kerr solution can be obtained in the limit $l\to\infty$, which is
the limit of  vanishing  cosmological constant. 
The horizon is defined by the largest root of the quartic algebraic
equation  
\beq
\Delta_r=0\,, \hs (r_H^2+a^2)(1+\frac{r_H^2}{l^2})-2mr_H=0\,.
\label{kh}
\eeq
Note that the solution is well defined only for $a^2<l^2$. In the
critical limit $a^2=l^2$, the three dimensional Einstein universe at
infinity rotates with the velocity of light. When the horizon angular
velocity $\Om<1/l$, a global
time-like Killing field can be defined outside the event horizon,
corresponding to the absence of super-radiance emission. 
The physical mass of the black hole relative to our choice of time
coordinate\footnote{One can normalize the Killing vectors so that the
corresponding charges generate the $SO(2,3)$ algebra at infinity.} 
 is $M=m/\Xi$, whereas $J=Ma/\Xi$ is the angular momentum
 \cite{Caldarelli:1998hg,Hawking:1998kw}. 

Let us consider the metric in the near-horizon approximation, since we have 
seen that it is sufficient to treat the leading effect of Hawking radiation.
One has (see, for example \cite{lecca}) 
\beq
ds^2&=&-A'(r_H,\theta)(r-r_H)dt^2+\frac{dr^2}{B'(r_H,\theta)(r-r_H)}+ 
\frac{\rho^2(r_H,\theta)}{\Delta_\theta} d\theta^2 \nn \\
&+&
\frac{\Sigma^2(r_H,\theta) \sin^2 \theta}{\rho^2(r_H,\theta)\Xi^2}
  d\chi^2\,,
\label{lecca}
\eeq 
where
\beq
\chi= \phi-\Omega  t\,.
\eeq
We also may consider the trajectories with $\theta$ and $\chi$
constants. Actually, it is a well known property of the Kerr black hole,
shared by the KAdS solution, that along geodesics in surfaces
$\theta=\theta_0$ constant, the combination $\phi-\Om t$ is finite on
the horizon, while both $\phi$ and $t$ diverge. 
As for the BTZ black hole, we may apply again the result of  Section 2 by 
writing
$I=-Et+J\phi+W(r,\theta)$, then passing to the variable $\chi$ (which
transforms $E$ into $E-\Omega J$) we arrive at 
\beq
\Gamma \equiv e^{-2 \mbox{Im}I}= \exp\left(-\frac{4\pi(E-\Omega J)}
{\sqrt{A'(r_H,\theta_0)B'(r_H,\theta_0)}}\right)\,.
\label{cckk}
\eeq
where $E-\Omega J>0$ must be assumed. This can be easily understood as
follows: the energy and angular momentum of a particle with
four-momentum $p^a$ are $E=-p^aK_a$ and $J=p^a\tilde{K}_a$,
respectively, where $K=\partial_t$ and $\tilde{K}=\partial_{\phi}$ is
the rotational Killing field. But the Killing field which is time-like
everywhere (including the ergosphere) is not $K^a$, but is instead
$\chi=K+\Omega\tilde{K}$. Hence a particle (including those
with negative energy inside the ergosphere) can escape to infinity
if and only if $p_a\chi^a<0$, which gives the wanted inequality

\[
p^a(K_a+\Omega\tilde{K}_a)=-E+\Omega J<0
\]

At the same time, it is violated only in
the super-radiant regime, where the Boltzmann distribution must be
replaced with the full Planck distribution, and thus it is outside the
reach of the tunneling semi-classical method.

The dependence on the constant angle $\theta_0$ is only apparent, because
one finds
\beq
A'(r_H,\theta_0)B'(r_H,\theta_0)=\frac{r^2_H}{(r_H^2+a^2)^2}
\left(1+\frac{3r_H^2}{l^2}+\frac{a^2}{l^2}-\frac{a^2}{r_H^2}\right)^2\,. 
\eeq
in agreement with the zeroth law, according to which the surface gravity
must be constant all over the horizon. As a result, rearranging a bit,
the Hawking temperature is 
\beq
T_H=\frac{3r_H^4+(l^2+a^2)r_H^2-a^2l^2}{4\pi lr_H(r_H^2+a^2)}. 
\eeq
and one can see that $2\pi T_H$ is indeed the surface gravity of the
black hole. 
Here we may note that the tunneling method, having the very
nature of a semi-classical scheme, only capture the
Boltzmann tail of the Hawking asymptotic
flux. In particular, the method misses the super-radiant
emission usually associated to rotating non extremal and charged black
holes. In this last case this may seem a little bit disappointing, since it
is well known \cite{Gibbons,Page,Vanzo} that super-radiance is driven by a
Schwinger process of pair production, amenable in principle to
semi-classical methods. Remarkably, it may not always be the case
that Schwinger pair creation takes place through tunneling, as has
been shown by Friedmann and Verlinde \cite{Friedmann:2002gx} in their
study of pair production of Kaluza-Klein particles in a static KK
electric field.

We conclude this Section by the following remark. The results obtained within
the tunneling method give rise again to Hawking temperature expressions in 
agreement
with the Euclidean one, according to which there should be no conical
singularity in the associated Euclidean, Wick rotated solutions.

\section{Conclusions}

In this paper, we have revisited the so called tunneling method, namely 
a simple semi-classical method useful in investigating the Hawking radiation 
issue. The method has been reformulated in the case of an arbitrary
static black  
hole solution  and restricted only to the leading term, namely neglecting the 
back reaction on the black hole geometry. The different role of
ingoing particles has been noticed. Then it has been extended to 
extremal black holes. In these cases, our 
recipe, consisting in the covariant treatment of the horizon singularity,
through the use of spatial proper distance, has allowed to derive the correct
result of zero Hawking temperature, but only for those solutions
having the horizon at infinite spatial distance. With  regard to this
issue, we have also  
stressed that the Euclidean method, based on the regularity
requirement of absence of  
conical singularities, works only in the non extremal case, and in this case, 
the results obtained within
the tunneling method give rise again to Hawking temperature expressions in 
agreement with the Euclidean one. The method has also been applied to the 
multi-horizons case, which is conceptually more difficult to interpret,
but the usually accepted Hawking temperatures have been recovered. 

We also have considered the stationary (rotating) case. Here we have 
investigated two important cases, namely the rotating BTZ black hole and 
the Kerr-AdS black hole solution. Again, making use of the near horizon 
approximation, the tunneling rate in the leading approximation has been 
derived and, as a consequence, an expression of the Hawking temperatures
agreeing with the values computed by means of standard methods. 

Finally, it should be noted that the fundamental thermal nature of
the radiation flux remains elusive, and is really inferred from the
tunneling method because the coefficient of energy is related to the
known temperature of the black hole in precisely the way required by
the Boltzmann distribution. A proper treatment should consider also the
absorption probabilities, and a check as whether detailed balance
between absorption and emission is really satisfied, before inferring
a Planck emission spectrum. That this is so, is of course a well known
piece of knowledge.

\end{document}